# Evaluating the Surface Tension Using Grand Canonical Transition Matrix Monte Carlo and Finite-Size Scaling

by

Jeffrey R. Errington[*]

*Department of Chemical Engineering, University at Buffalo, The State University of New York,*

*Buffalo, NY 14260-4200*

## Abstract

This letter describes a novel approach for determining the surface tension of a model system that is applicable over the entire liquid-vapor coexistence region. At the heart of the method is a new technique for determining coexistence properties that utilizes transition probabilities of attempted Monte Carlo moves during a grand canonical simulation. Finite-size scaling techniques are implemented to determine the infinite system surface tension from a series of finite-size simulations. To demonstrate the new method, the surface tension of the Lennard-Jones fluid is determined at temperatures ranging from the triple point to the critical point.

---

[*] Electronic address: jerring@buffalo.edu

The interfacial tension of fluids plays a key role in many industrial processes and scientific phenomena. Examples include the development of protective coatings and the self-assembly of amphiphilic polymers. Although the scientific and industrial importance of the interfacial tension has been realized for some time, development of a robust method for calculating the surface tension of a model system using molecular simulation techniques that is applicable over the entire liquid-vapor coexistence region has proven to be particularly challenging. There are two computational techniques that are commonly used for determining interfacial tensions. In one, the surface tension is calculated using elements of the pressure tensor evaluated during a molecular dynamics or Monte Carlo simulation in which an explicit interface is formed [1-7]. This approach is most appropriate at moderate to low reduced temperatures. As the critical temperature is approached, it becomes difficult to maintain the interface. Moreover, the results are known to be highly sensitive to the procedure used for truncating the potential and the size of the system simulated [8,9]. A second technique utilizes the finite-size scaling formalism introduced by Binder [10]. In this approach, grand canonical [11] or isothermal-isobaric [12] simulations are used to evaluate the free energy barrier between the liquid and vapor phase for a series of system sizes, after which finite-size scaling is used to extrapolate the infinite system value. This method has enjoyed success for calculations performed at temperatures close to the critical point. However, at lower temperatures, the calculations necessary to negotiate the liquid-vapor free energy barrier become intractable.

This paper describes a novel approach for directly determining the liquid-vapor phase coexistence, including the free energy barriers, of a model system at any temperature along the coexistence line. The method relies on transition matrix Monte Carlo ideas [13-17]. In particular, Monte Carlo transition dynamics and variance reduction techniques, developed by Fitzgerald *et* *al.* [16,17], are implemented. The transition matrix algorithms were originally developed for investigation of discrete systems. Here, the concepts are incorporated with the grand canonical ensemble to provide a framework for determining the thermodynamic properties of continuous systems. The new method utilizes information from the attempted *transitions* between states along the Markov chain as opposed to tracking the number of times the chain *visits* a given state. The data collection scheme for the proposed technique is highly efficient; at no point is any of the data discarded. Determination of surface tensions is completed using the finite-size scaling formalism of Binder [10]. This approach provides a means to estimate the behavior of the infinite system from finite-size calculations. The methods presented in this paper are general. To illustrate the concepts, the surface tension of the Lennard-Jones fluid is determined at temperatures ranging from slightly above the triple point to slightly below the critical point.

Consider a system in the grand canonical ensemble [18], where the volume *V*, temperature *T*, and chemical potential $\mu$ are fixed and the number of particles *N* and energy *E* fluctuate. For such a system, the Markov chain visits a given *microstate* $s = s(\mathbf{r}_1, \mathbf{r}_2, \ldots, \mathbf{r}_N)$ with probability,

$$\pi_s = \frac{1}{\Xi} \frac{V^{N_s}}{\Lambda^{3N_s} N_s!} \exp(-\beta E_s) \exp(\beta \mu N_s), \tag{1}$$

where $\beta$ denotes the inverse temperature ($\beta = 1/k_B T$, $k_B$ is the Boltzmann factor), $\Lambda$ is the de Broglie wavelength, and $\Xi$ is the partition function, defined such that $\sum_s \pi_s = 1$. For phase coexistence calculations, one must determine the probability $\Pi_N$ of finding the system with the *macrostate* variable *N*, given by $\Pi_N = \sum_{s, N_s = N} \pi_s$. To obtain an estimate of $\Pi_N$, a standard grand canonical simulation is performed with a simple bookkeeping scheme that enables one to deduce the transition probabilities $P_{N,N'}$, which indicate the probability of moving to state $N'$ given that the current state is $N$. The algorithm consists of three parts [17]:

1. For each Monte Carlo step, propose a move to a new microstate $s'$ from the current microstate $s$ with probability $q_{s,s'}$ [19].

2. Accept the proposed configuration with probability $a_{s,s'} = \min\{1, \pi_{s'}/\pi_s\}$.

3. Regardless of whether the move is accepted, update a matrix $C_{N,N'}$ as follows,

$$C_{N,N'} = C_{N,N'} + a_{s,s'} \quad \text{and} \tag{2a}$$

$$C_{N,N} = C_{N,N} + 1 - a_{s,s'} \tag{2b}$$

This bookkeeping scheme is superior to earlier techniques that required one to update the $C_{N,N'}$ matrix by either zero or one after each step [13]. At any point during the simulation, an estimate of the transition probabilities is obtained as follows,

$$\tilde{P}_{N,N'} = \frac{C_{N,N'}}{\sum_{N''} C_{N,N''}}, \tag{3}$$

where the tilde (~) is used to indicate an *estimate* of the transition probabilities. After obtaining an estimate of $P_{N,N'}$, the Monte Carlo detailed balance expression is employed to find the macrostate probabilities $\Pi_N$,

$$\Pi_N P_{N,N'} = \Pi_{N'} P_{N',N}. \tag{4}$$

In the above expression, the transition matrix $P_{N,N'}$ is a stochastic matrix with a stationary solution of $\Pi_N$. For the grand canonical simulation with a single particle creation or annihilation proposal mechanism, the transition matrix $P_{N,N'}$ is banded with three bands. This greatly simpli-fies the determination of $\Pi_N$. An iterative scheme can be used to determine the macrostate probabilities,

$$\ln \tilde{\Pi}_{N+1} = \ln \tilde{\Pi}_N + \ln \left( \tilde{P}_{N,N+1} / \tilde{P}_{N+1,N} \right). \tag{5}$$

To traverse the region between coexisting liquid and vapor phases, the system has to visit low probability states that are often inadequately sampled during a conventional grand canonical simulation. To address this problem, a simulation is often biased such that all macrostates within a specified interval are sampled with equal probability. This so called multicanonical sampling [20] is accomplished by assigning each macrostate a weight $\eta_N$ that is inversely proportional to the frequency of observing the macrostate during a conventional simulation, $\tilde{\eta}_N = -\ln \tilde{\Pi}_N$. The acceptance probability in step 2 above now becomes, $a_{s,s'}^{\eta} = \min\{1, e^{\eta_{N'}} \pi_{s'} / e^{\eta_N} \pi_s\}$. During the simulation, the weighting function is reevaluated at regular intervals. The success of the current approach lies in the efficiency of the technique. Although a bias in the sampling has been introduced, the update scheme for $C_{N,N'}$ remains the same, i.e., after each Monte Carlo trial, $C_{N,N'}$ is incremented by the probability that the proposed move would be accepted in the *absence* of multicanonical sampling. This simplification allows one to periodically update the weighting function without having to eliminate the previous data. Over time, the weighting function evolves and eventually the entire macrostate interval of interest is sampled. The run is terminated when the weighting function, and thus the particle number probability distribution, converges.

Once the macrostate probabilities have been determined, the coexistence properties are calculated in a straightforward manner. Histogram reweighting [21] is used to locate the chemical potential that provides coexistence at the given temperature. The coexisting densities are subsequently determined from the first moment of the corresponding liquid and vapor peaks of the $\Pi_N^{coex}$ distribution. If the zero particle limit is sampled during the simulation, the vapor pressure is determined using the ideal gas as a reference state [22,23],

$$\beta pV = \ln\left(\sum_N \Pi_N^{coex} \Big/ \Pi_0^{coex}\right) - \ln(2). \tag{6}$$

Finally, the liquid-vapor free energy barrier $F$ for a system with box length $L$ [24] is obtained from the maximum probabilities in the liquid $\Pi_{Nmax}^{liq}$ and vapor $\Pi_{Nmax}^{vap}$ phases and the minimum probability $\Pi_{Nmin}$ in the region enclosed by the two maxima [25],

$$\beta F_L = \frac{1}{2}\left(\ln \Pi_{Nmax}^{liq} + \ln \Pi_{Nmax}^{vap}\right) - \ln \Pi_{Nmin}. \tag{7}$$

The finite-size scaling formalism of Binder [10] is used to determine the liquid-vapor surface tension. According to this method, the finite-size interfacial tension $\gamma_L$ of a three dimensional system is given by,

$$\beta \gamma_L = \frac{\beta F_L}{2L^2} = c_1 \frac{1}{L^2} + c_2 \frac{\ln L}{L^2} + \beta \gamma, \tag{8}$$

where $\gamma$ is the infinite system interfacial tension and $c_1$ and $c_2$ are constants. The expression suggests that the term $\beta F_L/2L^2$ becomes linear in the scaling variable $\ln(L)/L^2$ as the system size approaches infinity. The formalism enables one to extrapolate the infinite system size interfacial tension from a series of finite system calculations.

The Lennard-Jones fluid [26] has been selected to demonstrate the new approach. The energy of interaction $u$ between any two particles in the system separated by a distance $r$ is given by,

$$u(r) = 4\varepsilon\left[(\sigma/r)^{12} - (\sigma/r)^6\right], \tag{9}$$

where $\varepsilon$ and $\sigma$ are energy and size parameters, respectively. From this point forward, all quantities are nondimensionalized using $\varepsilon$ and $\sigma$ as characteristic energy and length scales, respectively. For example, temperature is reduced by $\varepsilon/k_B$ and the interfacial tension by $\varepsilon/\sigma^2$. The method outlined above is used to evaluate the liquid-vapor interfacial tension at temperatures of $T = 0.70$, $T = 0.85$, $T = 1.10$, and $T = 1.30$. Grand canonical transition matrix Monte Carlo simulations are carried out with systems ranging in size from $L = 6$ to $L = 14$ using as many as 20 billion MC steps with the weighting function updated every one million MC steps. The potential cutoff is set at one-half the box length and standard tail corrections are applied [27].

The density distributions at coexistence for various system sizes with $T = 0.85$ are displayed in Figure 1. The results indicate that probabilities as low as $10^{-200}$ are accessible using the transition matrix approach. This range of probabilities is significantly greater than possible when using multicanonical methods that implement a visited states strategy [11].

Figure 2 shows the liquid-vapor interfacial tension as a function of the scaling variable for the four temperatures examined. One of the more striking features of the plot is the precision of the calculations. For example, at $T = 0.85$ the uncertainty in the interfacial tension is less than 0.5 percent of the mean for all system sizes [28]. As expected, the data points do not collapse onto a straight line as a function of the scaling variable, but rather display a small degree of curvature. The best method for extrapolating the infinite system interfacial tension from the finite-size data is not clear [12]. In this work, the infinite system interfacial tension was estimated from a straight line fit using data points with $L \geq 9$. Note that the infinite system values become more sensitive to the extrapolation technique as the temperature increases. For example, at $T = 0.70$ the $L = 12$ interfacial tension is within 3.9 percent of the infinite system result, whereas the equivalent value at $T = 1.30$ is 42 percent. The infinite system surface tensions and the corresponding coexistence properties are collected in Table 1.

The surface tensions obtained in this study, along with those from previous investigations, are displayed as a function of temperature in Figure 3. The agreement among the data sets is excellent. At low temperatures, the results generated using the transition matrix approach are consistent with the investigations of Mecke *et al.* [5] and Chen *et al.* [7], where techniques that require an explicit interface were employed. The approach presented here has a number of advantages over explicit interface methods: (i) the new technique is straightforward to use, requiring just slightly more overhead to implement than a standard grand canonical simulation, (ii) the precision of the results from this work appears to be superior to that of data from previous investigations, (iii) there exists a well-developed framework to describe how the interfacial tension scales as a function of system size, and (iv) the new method does not fail at near-critical temperatures. As expected, the current results are also consistent with those of Potoff and Panagiotopoulos [11], who utilized grand canonical ensemble simulations coupled with the visited states technique for obtaining multicanonical weights. The limitations of the visited states method prevented the authors from obtaining surface tensions below $T = 0.95$. The current study demonstrates that adopting the transition matrix approach allows one to probe the entire liquid-vapor coexistence region.

In summary, a general method has been presented for directly locating the phase coexistence of a model system at a given temperature. Implementation of the method requires the addition of a simple bookkeeping scheme that tracks the transition probabilities for addition and deletion of particles during a grand canonical simulation. The method is applicable over the entire liquid-vapor coexistence region and does not appear to be limited by system size. It is straightforward to couple the new technique to the finite-size scaling framework of Binder for accurate determination of interfacial tensions.

The outlook for the approach presented in this paper is bright. The results from this study indicate, that for a single component system containing spherically symmetric particles, the new method is just as straightforward to implement and potentially more accurate than equivalent Gibbs ensemble [29] calculations. Whether this conclusion holds true for more complicated systems will be addressed in future investigations. The overall performance of transition matrix methods has been remarkable. In addition to the current study, they have been used with equal success to determine solid-liquid phase coexistence and free energies of solvation of nonpolar molecules in water [30].

## Acknowledgement

I thank Professor David Kofke for helpful discussions and comments on the manuscript. Financial support for this project was provided by startup funds from the University at Buffalo. Computational resources were provided in part by the University at Buffalo Center for Computational Research.

# Tables

**Table 1.** Coexistence properties of the Lennard-Jones fluid.

| $T$ | $p^a$ | $\rho_{liq}^a$ | $\rho_{vap}^a$ | $\gamma$ |
|---|---|---|---|---|
| 0.70 | 0.001367 (2) | 0.8424 (13) | 0.001992 (2) | 1.182 (10) |
| 0.85 | 0.007618 (5) | 0.7760 (10) | 0.009611 (9) | 0.837 (2) |
| 1.10 | 0.04592 (5) | 0.6410 (7) | 0.05485 (5) | 0.343 (2) |
| 1.30 | 0.1212 (1) | 0.4271 (13) | 0.2096 (13) | 0.0050 (5) |

$^a$ Calculated using a system size of $V = 1000$.

# Figure Captions

**Figure 1.** The probability of observing a system with a given density while at coexistence with $T = 0.85$ for a range of system sizes.

**Figure 2.** The system size dependence of the effective surface tension. The plots from top to bottom are for $T = 1.30$, $T = 1.10$, $T = 0.85$, and $T = 0.70$. The circles represent simulation data and the dashed lines provide an extrapolation to infinite system size using data from systems with $L \geq 9$.

**Figure 3.** The surface tension of the Lennard-Jones fluid as a function of temperature. This work (filled circles), Mecke *et al.* [5] (squares), Chen *et al.* [7] (triangles), and Potoff *et al*. [11] (diamonds). The dashed line serves as a guide to the eye.

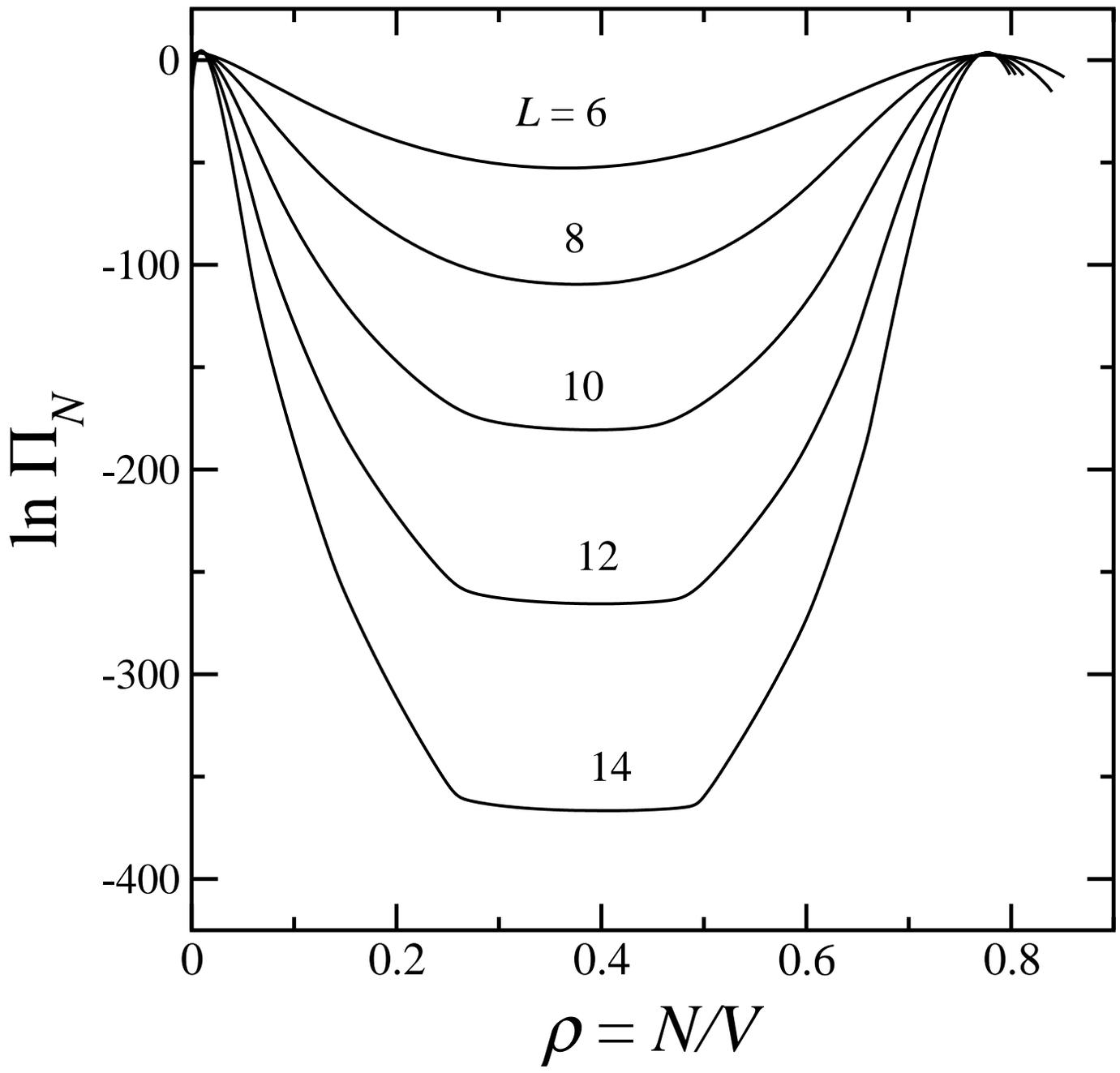

Figure 1

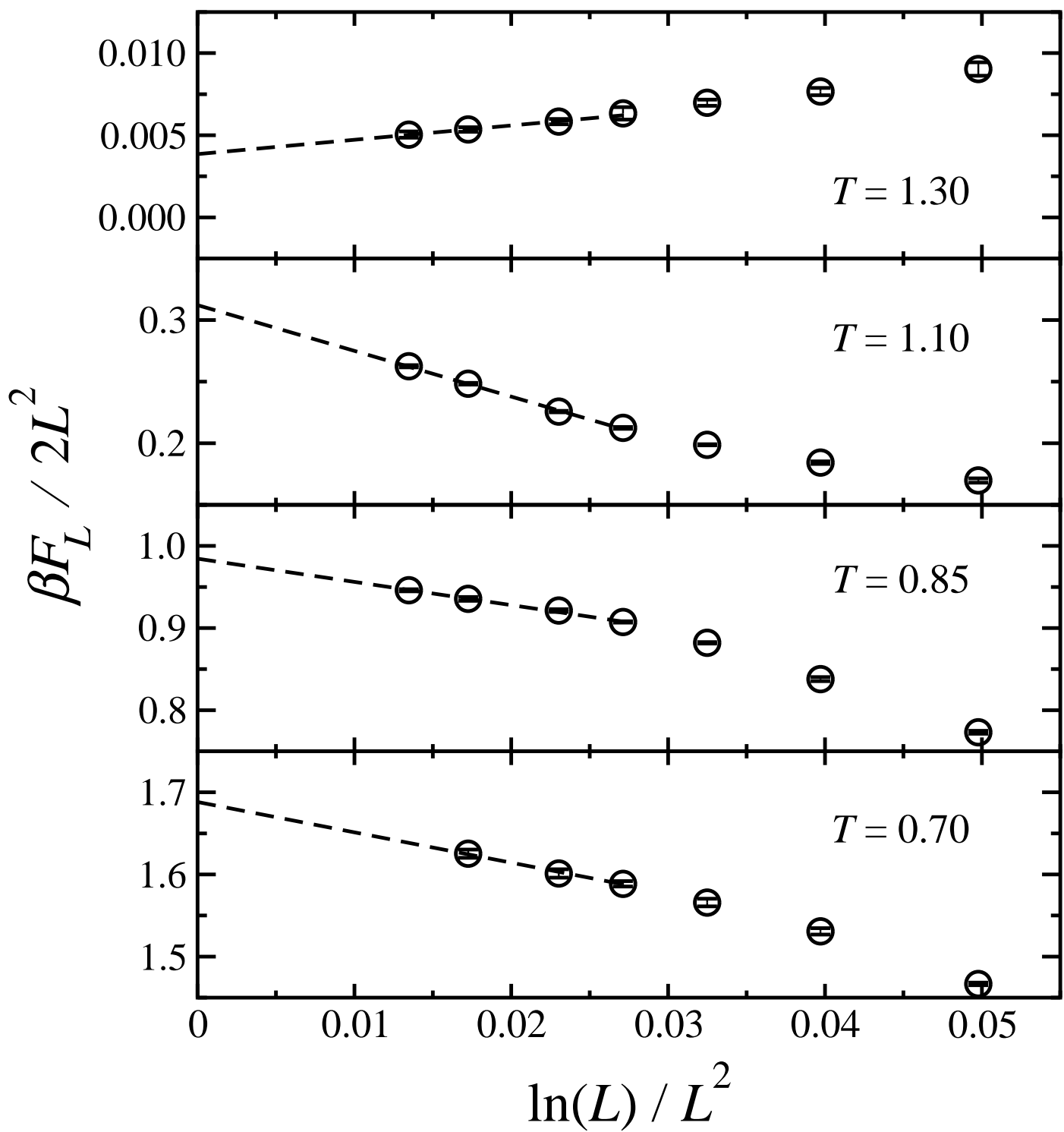

Figure 2

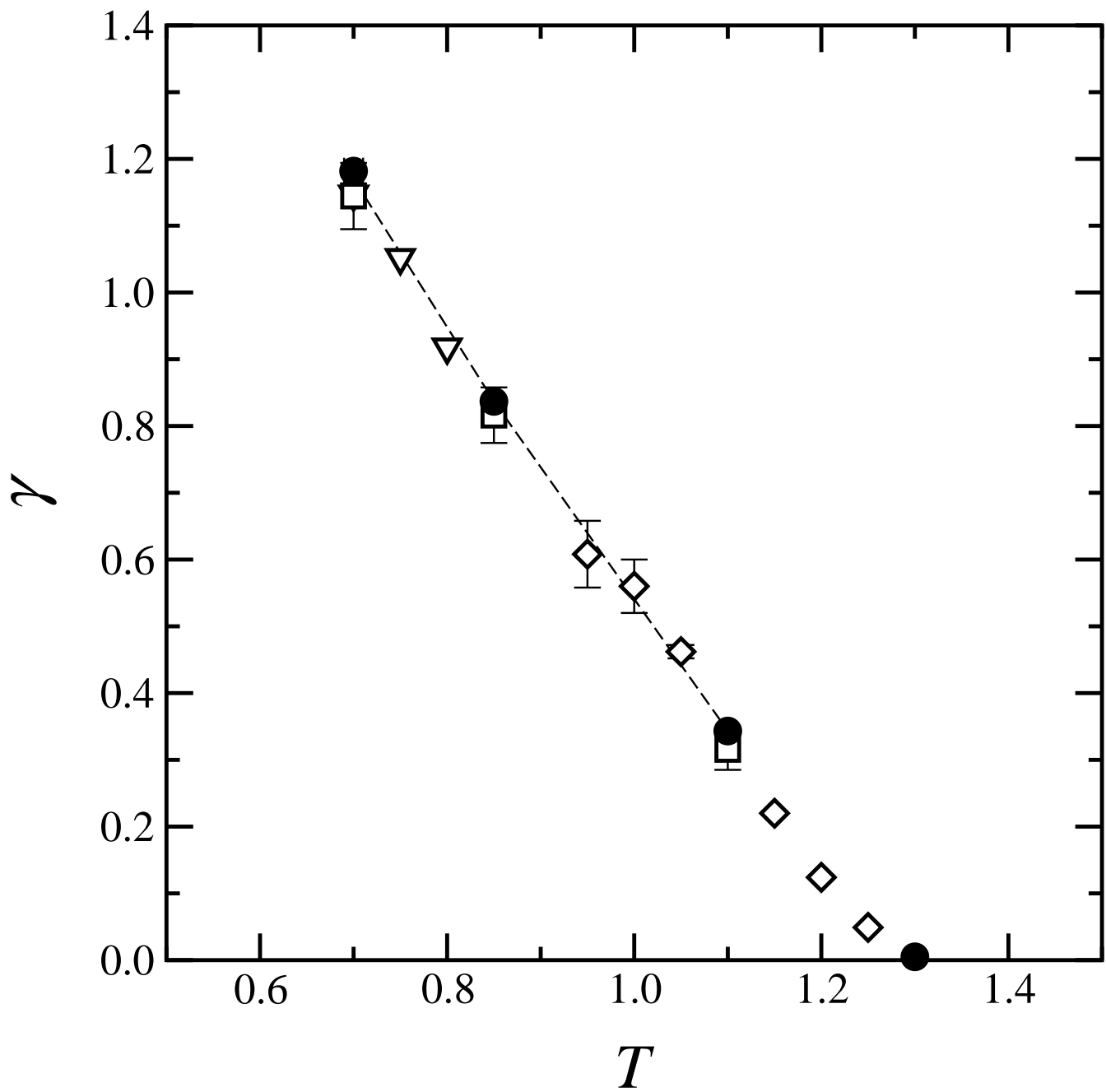

Figure 3